# Strain Modulated Superlattices in Graphene

**Authors:** Riju Banerjee[1]*, Viet-Hung Nguyen[2], Tomotaroh Granzier-Nakajima[1], Lavish Pabbi[1], Aurelien Lherbier[2], Anna Ruth Binion[1], Jean-Christophe Charlier[2], Mauricio Terrones[1], Eric William Hudson[1]*

**Affiliations:**

[1]Department of Physics, The Pennsylvania State University, University Park, PA 16802, USA

[2]Institute of Condensed Matter and Nanosciences, Université catholique de Louvain, Chemin des étoiles 8, B-1348 Louvain-la-Neuve, Belgium

*Correspondence to: ehudson@psu.edu, riju@psu.edu

**Abstract:** Strain engineering of graphene takes advantage of one of the most dramatic responses of Dirac electrons enabling their manipulation via strain-induced pseudo-magnetic fields. Numerous theoretically proposed devices, such as resonant cavities and valley filters, as well as novel phenomena, such as snake states, could potentially be enabled via this effect. These proposals, however, require strong, spatially oscillating magnetic fields while to date only the generation and effects of pseudo-gauge fields which vary at a length scale much larger than the magnetic length have been reported. Here we create a periodic pseudo-gauge field profile using periodic strain that varies at the length scale comparable to the magnetic length and study its effects on Dirac electrons. A periodic strain profile is achieved by pulling on graphene with extreme (>10%) strain and forming nanoscale ripples, akin to a plastic wrap pulled taut at its edges. Combining scanning tunneling microscopy and atomistic calculations, we find that spatially oscillating strain results in a new quantization different from the familiar Landau quantization observed in previous studies. We also find that graphene ripples are characterized by large variations in carbon-carbon bond length, directly impacting the electronic coupling between atoms, which within a single ripple can be as different as in two different materials. The result is a single graphene sheet that effectively acts as an electronic superlattice. Our results thus also establish a novel approach to synthesize an effective 2D lateral heterostructure – by periodic modulation of lattice strain.

**Main Text:**

Due to its high electronic mobility, optical transparency, mechanical strength and flexibility, graphene is attractive for electronic applications[1,2]. However, several factors prevent the realization of common electronic applications. For example, the lack of a band gap prevents an effective off-state in graphene transistors. Furthermore, Klein tunneling[3], in which electrons pass through an electrostatic barrier with perfect transmission, prevents electron confinement by traditional gating methods.

As an alternate means of electronic control, the effects of inhomogeneous magnetic fields on graphene have been theoretically investigated[4–12]. Configurations like square-well magnetic barriers, magnetic dots, and magnetic rings are all predicted to confine electrons in graphene[6,12]. In other cases, where the average B-field is zero, strong resonances that lead to wave vector-[7,8] and valley-[9] filtering are predicted. The physical manifestation of such inhomogeneous magnetic field configurations could lead to new graphene electronic, spintronic and valleytronic devices[4–10,12].

However, these theoretical proposals have remained unrealized because unlike engineering of strong electric field profiles with nanoscale variations, it is difficult to generate large magnetic fields that vary appreciably on the nanometer length scale. Using presently available techniques, strong magnetic fields created by large magnets are homogeneous at the length scale of most samples, while only weak inhomogeneous fields can be applied to a sample, for example by local magnetic strips. As an alternative approach, the application of strain gradients can lead to the emergence of strong pseudo-gauge fields which manipulate the electronic properties of graphene at the nanoscale[13]. Levy et al.[14] showed that pseudo-Landau levels can be observed in isolated graphene nanobubbles corresponding to locally uniform pseudo-magnetic fields of greater than 300 T. However, there the magnetic field strength (measured at the crest of the nanobubble) was relatively uniform and did not vary appreciably over the magnetic length of about 1.5 nm. Such isolated nanobubbles, thus, cannot be used to realize the theoretical proposals of novel electronic, spintronic and valleytronic devices requiring inhomogeneous fields[4–10,12]. Here, we demonstrate a novel technique to manipulate graphene's electronic properties by periodically modulating lattice strain, creating a superlattice of intense pseudo-gauge fields that oscillate with a spatial periodicity of a few nanometers, which is comparable to the magnetic length scale.

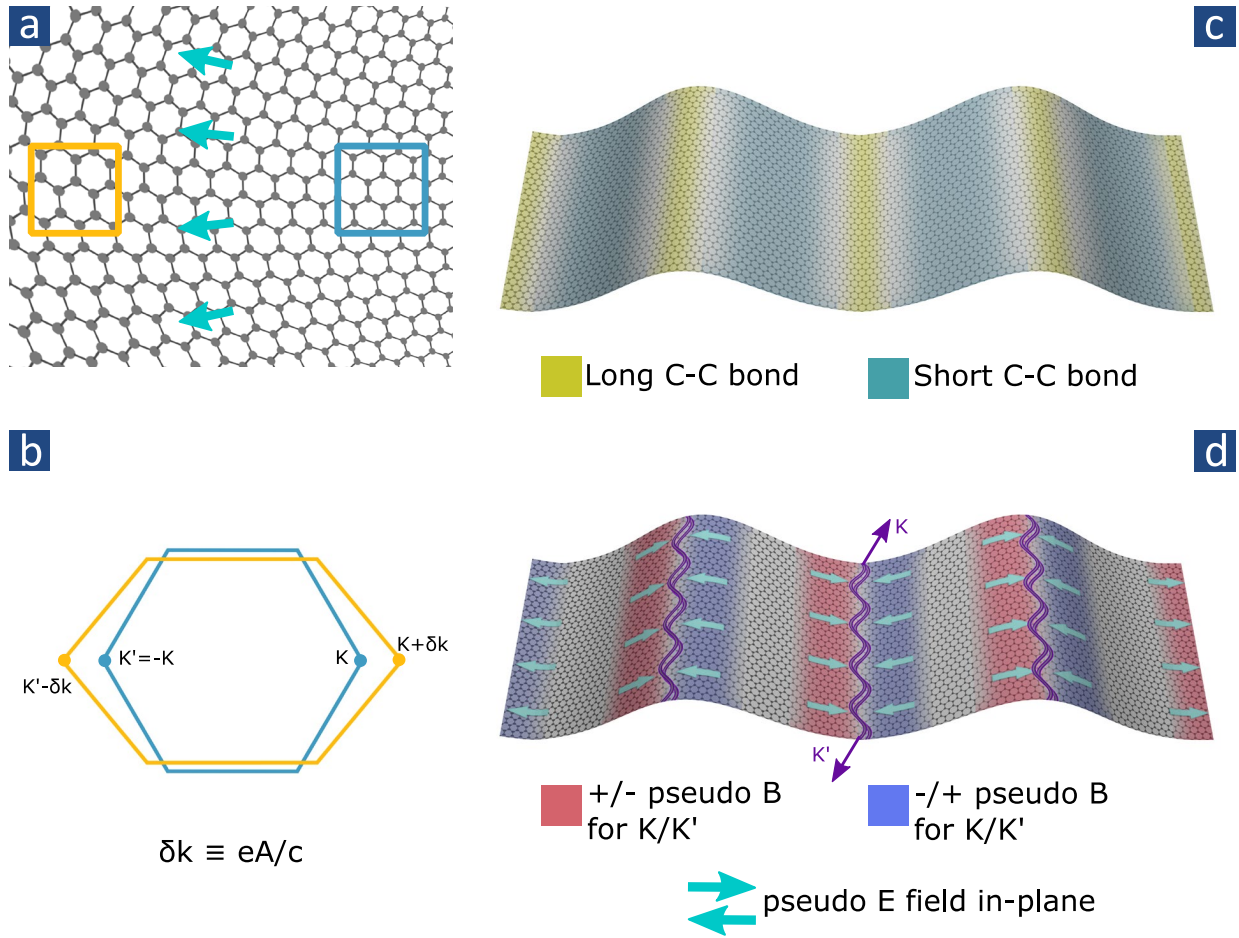

**Figure 1. Engineering periodic pseudo electric and magnetic fields at strained interfaces:** (a) High(low) density of carbon atoms and hence electrons are created in regions marked by light-blue (yellow) regions due to a strain gradient. This inhomogeneous charge distribution results in an electric field (green arrows). (b) Stretching of bonds cause the Dirac cones at K and K' points to shift symmetrically (yellow) from their original unstrained positions (light-blue) in the reciprocal space. As a momentum shift $\delta k$ can be interpreted as generating a pseudo-vector potential term $eA/c$[15], (where $e$ is the electronic charge and $c$ is the velocity of light) this creates pseudo-magnetic fields with opposite signs at the two valleys. (c) The strain associated with rippling creates rare (yellow) and dense (turquoise) regions in the graphene, effectively acting as two different materials in a superlattice. (d) Pseudo-fields form near the interfaces of these "materials," both electric (green arrows) and magnetic (red/blue regions indicating the $\pm\hat{z}$ field direction for pseudospin up electrons respectively; pseudospin down are flipped). The up and down magnetic fields are separated by only few nanometers, same order as the magnetic length, making the individual Landau levels to interact. LDOS peaks are maximized at the ripple crests and troughs, where valley polarized snake states (violet curved lines) are also expected to form due to the reversal of the pseudospin dependent pseudo-magnetic fields across these lines.

Pseudo-gauge potentials arise in graphene due to lattice strain[16–18] (Fig. 1). As argued by Suzuura and Ando[17], stretching the graphene lattice changes the local electron density which results in an in-plane electric field at the interface of regions with different strains (Fig. 1a). Strain also deforms the hexagonal structure in reciprocal space (Fig. 1b), moving the Dirac cones at the $K$ and $K'$ points in opposite directions. As changing the momentum $K \rightarrow K + \delta k$ can be interpreted as a pseudo vector potential[15], strain gradients also create a pseudo-magnetic field perpendicular to the graphene plane[16]. However, unlike an externally applied magnetic field which affects all electrons in a graphene lattice equivalently, the pseudo-magnetic field has opposite signs for the $K$ and $K'$ valleys. It has also been shown that the effects of any strain gradient in graphene can be modeled by pseudo-magnetic and electric fields[19–21]. It is straightforward to imagine that instead of a uniform strain gradient, which creates uniform pseudo-gauge fields[16–18], a spatially oscillating strain gradient profile can be used to create a spatially oscillating pseudo-gauge field profile required to realize the various novel electronic, spintronic and valleytronic devices discussed earlier[4–10,12]. In this article, we realize such a system by modulating the C-C bond length creating a superlattice of regions which are locally dense and rare (Fig. 1c). Alternating zones of oppositely directed pseudo-gauge fields arise at the interfaces of these regions (Fig. 1d). As the magnetic length in each up/down pseudo-magnetic field region (colored red and blue in Fig. 1d) is comparable to the separation between the two regions, Landau levels in the two regions interact resulting in a new quantization distinct from the familiar Landau quantization in uniform fields.

To modulate strain in a graphene lattice, and realize a pseudo-gauge field superlattice, we use low pressure chemical vapor deposition (LPCVD) to grow graphene on electropolished Cu foils at 1020°C (See Supplementary Sec. S1). Previous studies[22] revealed that such high temperatures result in the formation of large Cu steps separated by relatively flat terraces (Fig. 2a). We find that graphene sheets grown by this method form continuous films that are pinned on the flat terraces and drape over the large (up to ~35 nm high) step edges. Upon cooling the sample slowly to 80K, the draped graphene experiences tensile and shear stresses as it gets pulled by the contact forces of the terraces. This leads to a periodic arrays of ripples, creating a strain enabled modulated superlattice (STREMS), as imaged by a scanning tunneling microscope (STM) (Fig. 2a and b).

To demonstrate the emergence of a pseudo-gauge field superlattice in STREMS, we take differential conductance spectra, representative of the local electronic density of states (LDOS), away from and on top of the ripples (at points A and B in Fig. 2b respectively). As shown in

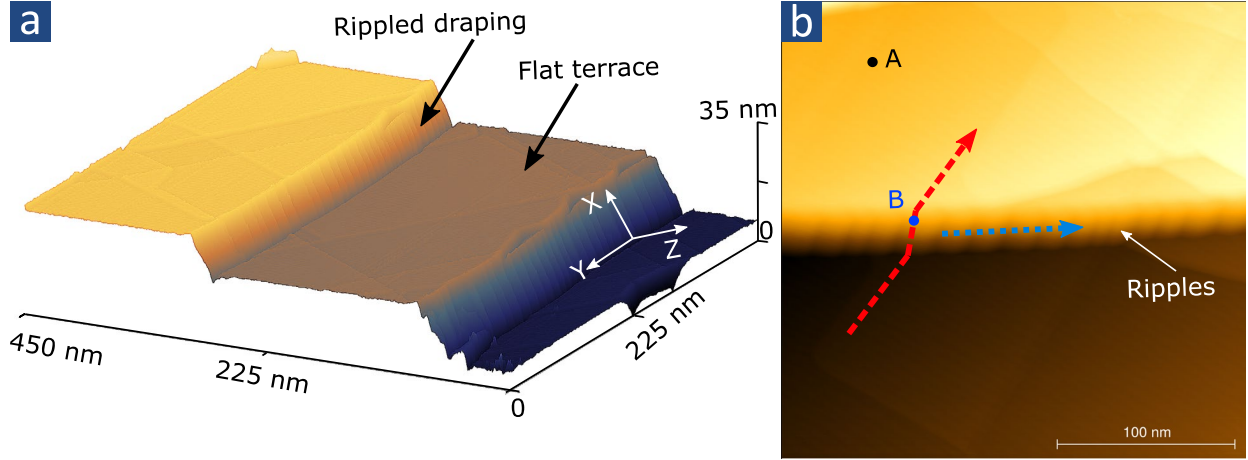

**Figure 2: Realizing Strain Modulated Superlattices:** (a) 3D-STM topography of 450 nm square region shows graphene draped over two steps separated by flat terraces. The draped graphene form ripples due to strain. Current setpoint $I_{set}$ = 62 pA, sample bias $V_s$ = 0.1 V. (b) STM topography zooming in on a step edge, highlighting locations at which data presented in later figures was obtained. $I_{set}$ = 80 pA, $V_s$ = 0.1 V. All topographic data is obtained at T = 80 K and unfiltered.

Fig. 3a, while spectra taken on the terraces (black curve) display the familiar V-shaped Dirac cone[23], those taken on STREMS (blue curve) reveal a series of peaks. Treating the terrace spectrum as a background, we subtract it to determine the strain-induced spectral modification (red curve). The fact that the strain-induced peaks are confined to the STREMS may be seen in a line-cut (series of spectra) across it (along the red arrow profile in Fig. 2b). In Figs. 3b and 3c we plot the height at which the spectra were obtained while climbing the step edge and the background-subtracted spectra respectively. It is clear from the figures that the LDOS peaks appear only in the spectra from the draped and rippled region (#8-19, colored red). Interestingly, these peaks are equally spaced in energy, scaling as $E_n \propto n$ (Fig. 3d). In addition to this energy dependence of the peaks, we also find a spatially-periodic modulation of their amplitude, as shown in a line-cut along the STREMS (Fig. 2b, blue arrow). In Fig. 4a we show that the amplitude of the LDOS peaks (red curve) is closely tied to the z-height profile (black curve), peaking at the ripple crests and troughs of the triangular ripples.

We note that these LDOS peaks are superficially similar to those found in a variety of previous STM measurements of locally strained (e.g. wrinkled or bubbled) graphene[14,20,24,25], where the LDOS is associated with Landau levels arising from strain-induced uniform pseudo-magnetic fields. However, unlike our observed linear scaling (implying $E_n \propto n$, in Fig. 3d), the energies of those peaks were reported[14,20,24,25] to scale as $E_n \propto \sqrt{n}$. Previous studies have observed equally spaced peaks in LDOS spectra in graphene and attributed them to confinement effects[26]. While it is possible to confine graphene electrons by strain[27,28], we discard that as a feasible explanation of our observed LDOS peaks as contrary to confinement peaks, the energy gap between our LDOS peaks remains unchanged for spectra taken on ripples with very different

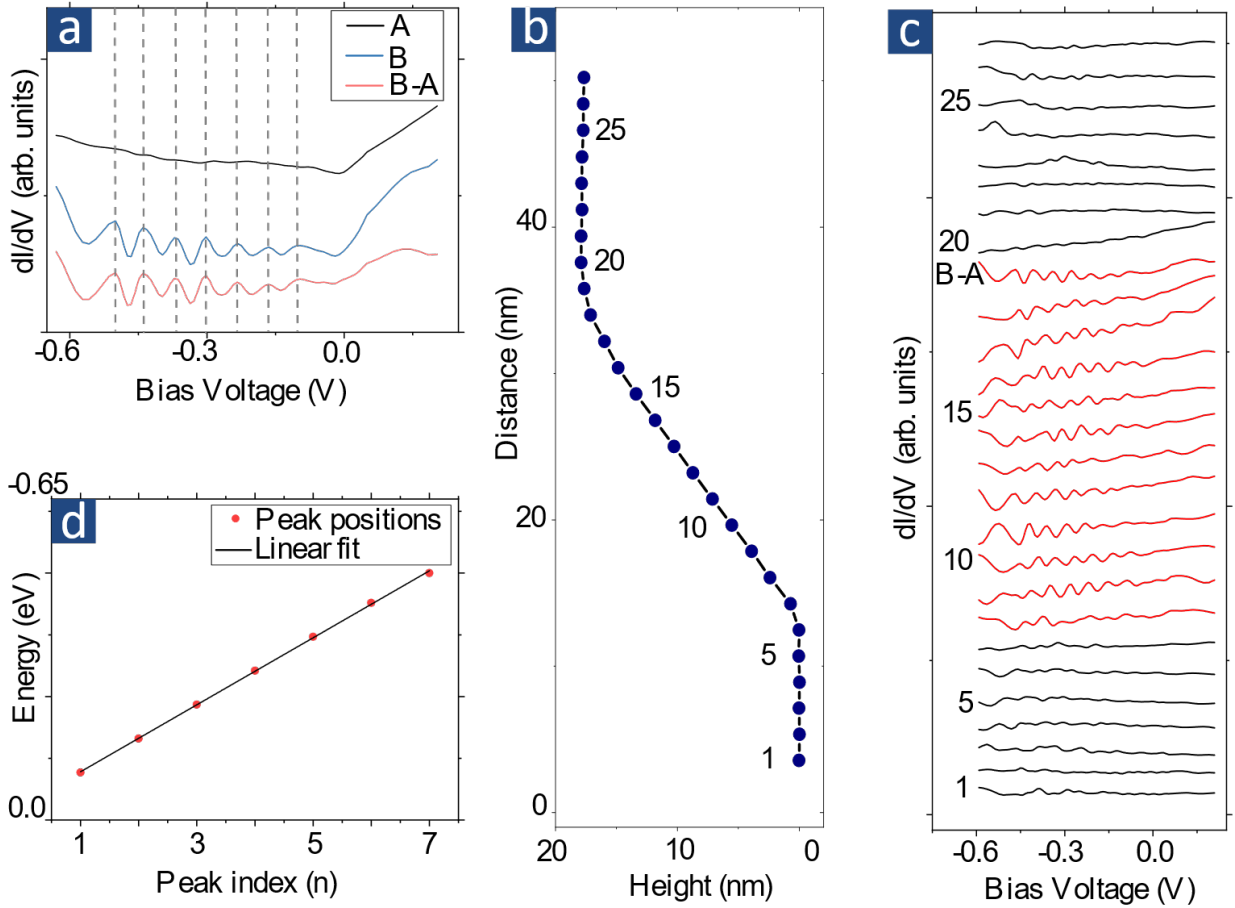

**Figure 3. Spectroscopy across a STREMS.** (a) Differential conductance spectra taken (A) away from and (B) on a STREMS (locations in Fig. 2b). Spectrum (A) has the Dirac point near the Fermi energy ($V = 0$ V). Treating (A) as background, subtraction highlights a series of peaks in the LDOS. Dashed lines aligned with the LDOS peaks are almost equally spaced and are guides to the eye. (b) A line cut rising across the STREMS (along red arrow of Fig. 2b) shows that peaks in the background subtracted spectra (c) only appear in the draped region (labelled #8-19, colored red). The Dirac cone shapes of the spectra is suppressed by subtracting the background from each, and the curves are offset for clarity. All these peaks are equally spaced, as shown in (d), plotted for the peaks in spectrum shown in (a), with slope = 67.83(5) meV/peak. For the twelve such spectra in (c), the average slope of energy vs. index plots is 69(3) meV/peak. See Supplementary sec. 6 for linear fits to all spectra taken on the draped region. All spectra are obtained using standard lock-in techniques, with 13mV bias modulation at ~971 Hz.

wavelengths (see supplementary sec. S2).

To clarify the origin of our LDOS peaks, atomistic calculations based on a $p_z$ tight-binding Hamiltonian with nearest neighbor couplings were performed with $H_{tb} = \sum_{\langle n,m \rangle} t_{nm}\, c_n^\dagger c_m$, where the hopping energies $t_{nm}$ are determined as a function of the C-C bond length $r_{nm}$ [29]:

$$t_{nm} = t_0 exp\left[-\beta\left(\frac{r_{nm}}{r_0} - 1\right)\right] \tag{1}$$

$t_0$ = -2.6 eV and $r_0$ = 1.42 Å were used as the corresponding parameters for unstrained graphene. The decay rate $\beta$ was adjusted to 4.5 by fitting first-principles calculations, while a value of 3.37 is usually used in flat graphene systems with in-plane deformations only[29]. Due to the exponential dependence of the hopping energy on the bond length $r_{nm}$, equation (1) implies that even a small change in $r_{nm}$ can have significant effects in modifying the local electronic properties. See Supplementary sec. S3 for more details about the calculations and the reason to require this modification in $\beta$.

To understand the profile of mechanical deformations, we consider a graphene sheet stretched under a constant tensile strain along the x-axis (axes directions are defined in Fig. 2a). A sinusoidal strain profile with out-of-plane displacements $h(y) = h_0 sin\left(2\pi\frac{y}{\lambda}\right)$, with $\lambda$ being the ripple wavelength, is then used to emulate the periodic ripples along the y-axis. In addition, curvature at the crests and troughs can also result in an in-plane displacement profile which can be modeled by a strain profile $u_y(y) = u_0 sin\left(4\pi\frac{y}{\lambda}\right)$. The two displacement profiles (out-of-plane and in-plane) are shown in Fig. 4b. The two different displacement profiles can be used to describe two significantly different modes of deformations which can occur in a ripple. For a given displacement profile $h(y) + u_y(y)$, when $u_0$ is sufficiently small (or zero), a strain profile is obtained with small C-C bond lengths at the ripple crests and troughs, and larger C-C bond lengths in the regions in between. However, for large enough $u_0$, the opposite situation occurs- describing a deformation profile with large C-C bond lengths at the crests and troughs and smaller C-C bond lengths in between. Although this simple model does not exactly reproduce the triangular shape of ripples observed experimentally, varying these two displacement fields allows us to investigate various deformation profiles and provides insights required to identify the model which agrees well with experimentally measured values.

Simulated LDOS obtained for the two situations discussed above (without and with in-plane displacements) are presented in Figs. 4c and d respectively. In both cases (Fig. 4c and d), we observe that a periodic strain profile described by $h(y)$ and $u_y(y)$ results in a series of almost equally spaced LDOS peaks, similar to our experimental observations (Fig. 3a, blue curve). The weakly varying peak spacing in the simulated LDOS (of ~80 meV) is also consistent with the experimentally observed peak spacing of 69(3) meV (in Fig. 3a, c and d); the small difference attributable to the fact that the simulation estimates the deformation profile as sinusoidal functions, instead of the observed triangular ripples. This is in stark contrast to a multitude of previous studies[14,20,24,25] where uniform strain gradients resulted in LDOS peaks corresponding to $E_n \propto \sqrt{n}$. More interestingly, while both models yield equally spaced LDOS peaks, the spatial variation of the amplitudes of those peaks is model dependent. In Fig. 4c, where we consider the out-of-plane displacement $h(y)$ only, higher amplitude peaks occur in the region between the crests and troughs of the ripples. In contrast, in Fig. 4d, where we consider both out-of-plane and in-plane displacement profiles ($h(y)$ and $u_y(y)$), higher amplitude peaks occur at the crests and troughs.

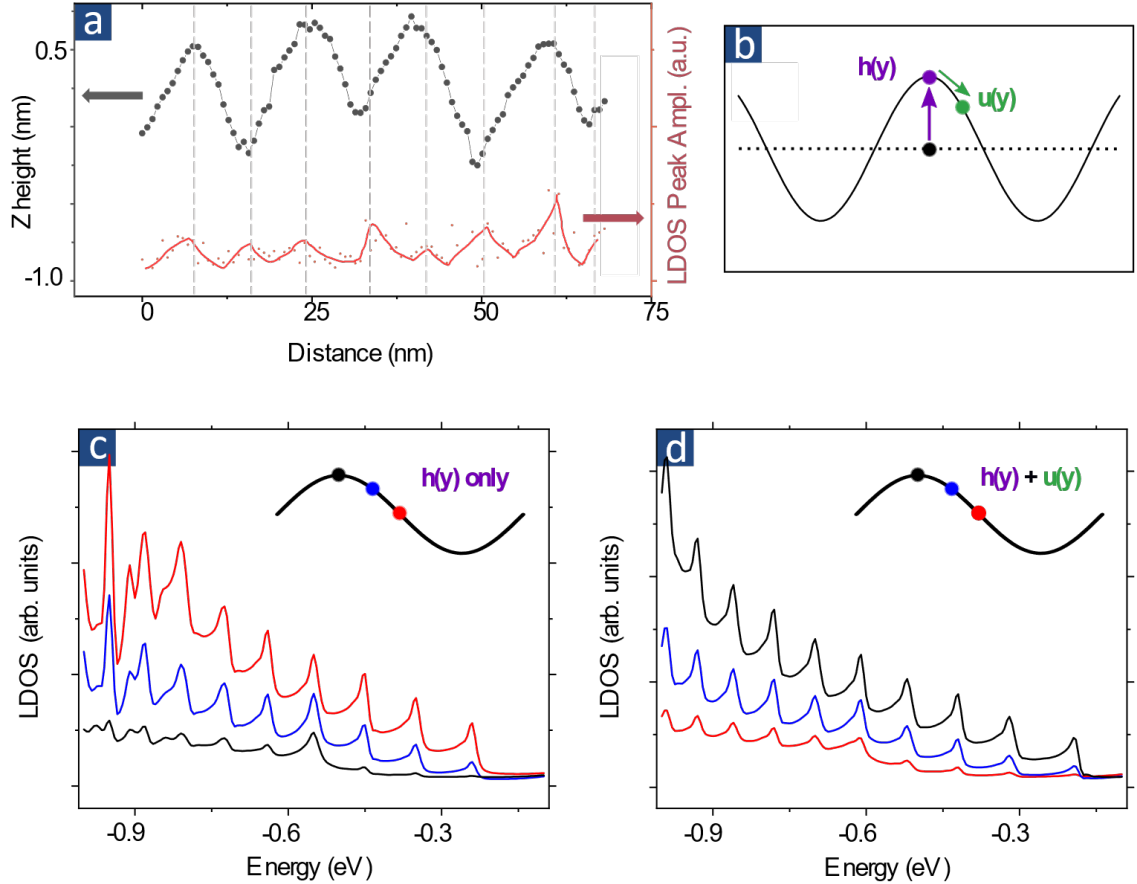

**Figure 4. Peak amplitude along the ripple.** (a) A line cut along the ripple (blue arrow in Fig. 2b) reveals that the amplitude of LDOS peaks (red) is maximized at the crests and troughs of the nearly triangular ripples (black). All lines are guides to the eye. (b) Displacement (out-of-plane $h(y)$ and in-plane $u(y)$) fields to be considered in rippled graphene. Simulated LDOS as a function of electron energy obtained when only out-of-plane displacements are considered (c) and when in-plane displacements are added (d) show that in-plane displacements are necessary to match the experimental results. Simulations calculated using displacement parameters $\lambda = 20$ nm, $h_0 = 2.5$ nm and $u_0 = 3.5$ Å.

As the latter (Fig. 4d) is consistent with our measurements (Fig. 4a), in-plane displacement must be included $(u_y(y) \neq 0)$ in our calculations, implying that both displacement fields $h(y) + u_y(y)$ concurrently occur and play a major role in determining the electronic properties of rippled graphene. Many theoretical studies have treated out-of-plane and in-plane deformation of graphene independently[28,30–33]. Fig. 4 refutes such an assumption as both deformations occur concurrently and are significant enough to modify the local electronic properties.

While the strain profile of ripples can be deciphered from measuring the local electronic properties and fitting it to a theoretical model, such a periodic oscillation of atomic displacements is very difficult to directly image using microscopy techniques. Though atoms can be imaged, extracting an accurate value of strain over a very narrow region at the crests and troughs by measuring individual atomic displacements of a few percent of the C-C bond length is challenging. So, to demonstrate that strain varies at the crests and troughs of the ripples, we took the indirect route of measuring resultant electronic properties ($E_n \propto n$ with higher LDOS weight at the crests and troughs) and fitting it to a theoretical model (detailed in Figure 4). The local electronic properties can be measured much more accurately (as each spectrum is averaged 150 times) and enables us to do a more thorough analysis than measuring atomic displacements.

Next, we turn to our observation of equally spaced ($E_n \propto n$) LDOS peaks (Fig. 3d) as opposed to $E_n \propto \sqrt{n}$ peaks in previous studies of strained graphene[14,20,24,25]. Under the effect of a spatially uniform magnetic field, Dirac electrons settle into Landau levels with $E_n \propto \sqrt{n}$ quantization. A crucial component of this familiar Landau quantization is the spatial uniformity of the magnetic field. The electronic quantization deviates from the familiar Landau quantization for spatially non-uniform magnetic field profiles. To illustrate this point, we show the DOS of a model system of Dirac electrons under a spatially periodic magnetic field profile with periodicity $L_B$ (Fig. 5a). In the limit $L_B \to \infty$, the DOS corresponds to standard Landau quantization with $E_n \propto \sqrt{n}$ (black curve in Fig. 5b). However, as the periodicity $L_B$ is reduced (blue curve in Fig. 5b, corresponding to $L_B = 34\ nm$), quantum Hall edge states are formed in each uniform B-field zones but propagate oppositely in zones of opposite B-fields. The interaction between these opposite propagating Hall states at the interface between zones of opposite B-fields modifies the quantization picture. In addition to the high DOS peaks corresponding to the standard Landau quantization, many smaller DOS peaks are observed in the blue curve in Fig. 5b. These small DOS peaks are essentially due to the aforementioned interaction between opposite propagating states that can induce the interference effects as investigated in a recent calculation by Nguyen and Charlier[34]. On reducing the periodicity further (red curve in Fig. 5b) and making $L_B$ the same order of magnitude as the magnetic length, the DOS profile is drastically different from the Landau quantized states. The red curve shows peaks which are almost equally spaced as observed in our experiments. Thus, we interpret our observation of $E_n \propto n$ LDOS peaks as a direct consequence of having a spatially oscillating magnetic field profile. Previous studies of strained graphene have reported the emergence of pseudo-gauge fields in structures like nanobubbles[14] and wrinkles[24,35,36] where the pseudo-magnetic field has been constant over a length scale much larger than the magnetic length . The fact that there are no experimental works yet detailing the results of inhomogeneous pseudo-magnetic fields is because prior to the present work, inhomogeneous field profiles have not been created with any regularity, to make the system amenable enough for thorough experimental and theoretical explorations. We also present an analytical derivation showing the emergence of $E_n \propto n$ quantization for Dirac electrons under a periodic magnetic field profile in Supplementary sec. S7. Note that the above analysis is independent of the orientation of the graphene sheet with respect to the direction of the periodicity of the magnetic field. Similar experimental observations like the ones reported were made on different step edges which curve significantly proving the fact that a pseudo-gauge field superlattice is formed by a periodic strain profile irrespective of the small details like orientation of the graphene with respect to the Cu substrate and periodicity of the magnetic field.

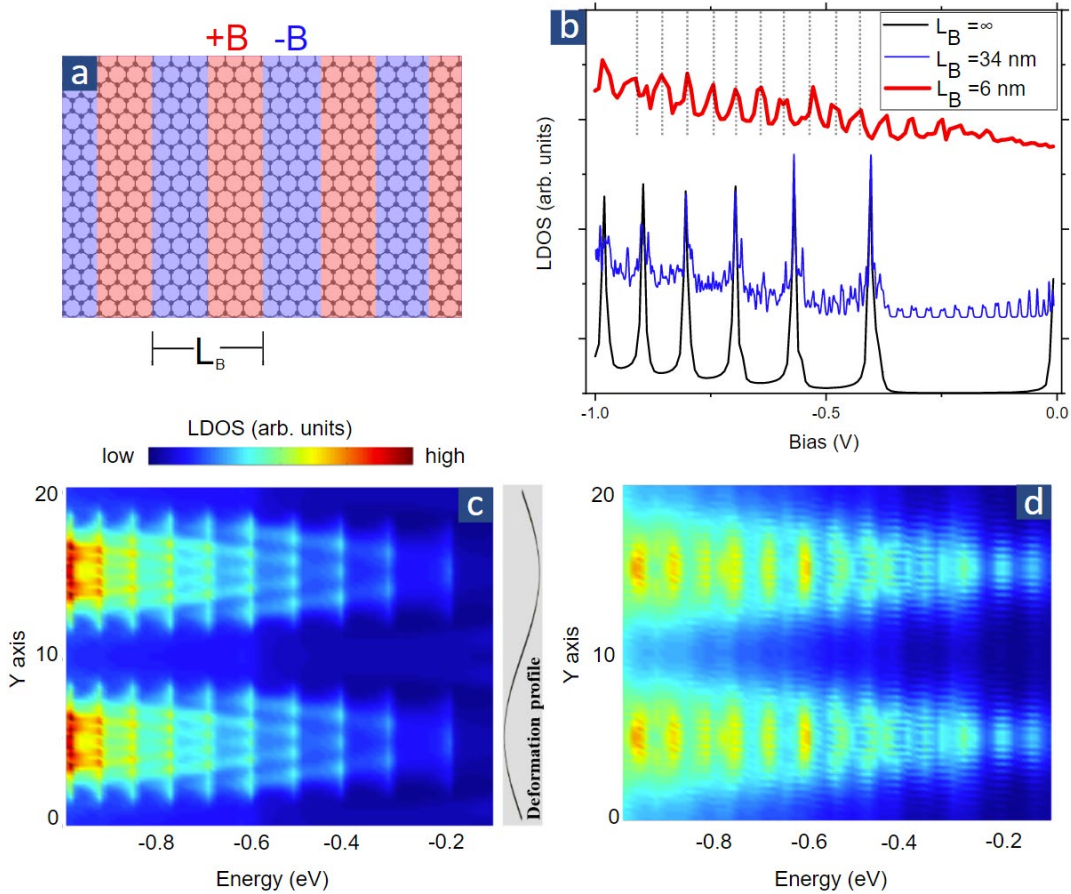

**Figure 5. DOS in periodic gauge fields:** (a) A model system of Dirac electrons in alternating magnetic field directions with periodicity $L_B$ is used to understand our experimental LDOS. (b) shows the calculated DOS for different values of $L_B$. The black curve, corresponding to $L_B \to \infty$ corresponds to the case of standard Landau quantization with energy peaks at $E_n \propto \sqrt{n}$. However, as the periodicity is reduced (blue curve), the spectra deviates from the standard Landau picture with more features. When the periodicity $L_B$ is reduced further and made of the same order of magnitude as the magnetic length, the DOS (red curve) is drastically different from the Landau quantized states (black curve). The peaks in the red curve are almost equally spaced, as observed in our experiments. All calculations in (b) are done with $|B| = 160$ T. The plots are offset for clarity. Equally spaced dashed lines are guides to the eye. (c) and (d): Theoretical calculations reproducing the observed spatial variation of these quantized states. Both (c) Strained and (d) unstrained + pseudopotential tight binding models capture important features in spectroscopy obtained along the ripple (blue arrow in Fig. 2b). LDOS peaks are almost equally spaced in energy, as observed in Fig. 3d. The LDOS amplitudes are higher at the crests and troughs of the ripples, as observed in Fig. 4a. The simulation also captures the feature of having higher LDOS weight at large negative biases, as observed in Fig. 3a.

A further demonstration of the emergence of spatially periodic pseudo-magnetic and electric fields in our rippled graphene system is presented in Fig. 5 c and d. We consider an effective model of unstrained flat graphene with a tight binding Hamiltonian, pseudomagnetic field

$\vec{B}_S = B_{max} \sin\left(4\pi \frac{y}{\lambda}\right) \vec{e}_z$ and electric potential $V_S = V_{max} \cos\left(4\pi \frac{y}{\lambda}\right)$, and compare it to a strained tight-binding model using the deformation profile discussed earlier (with both out-of-plane and in-plane deformations). LDOS maps from these two models are displayed in Figs. 5d and c respectively. The result obtained from an unstrained model with only including the combined effects of spatially varying E- and B- fields (Fig. 5d) is in good qualitative agreement with that obtained in the periodically strained system (Fig. 5c). This implies that both pseudo-magnetic and electric fields are induced by the corresponding strain field and play significant roles in influencing the electronic properties of the system. For a more detailed explanation for the requirement of pseudo-electric fields in addition to pseudo-magnetic fields, and to explain the spatial variation of the observed spectra, see supplementary section **S4**.

The above investigations confirm the generation of strain modulated superlattices, as displayed in Fig. 1c, d. We apply extreme strain to a graphene sheet, measuring over 10% as both estimated from our simulated models (Supplementary sec. 3) and measured directly by imaging the graphene lattice (Supplementary sec. 5). Analogous to a classical fabric under stress in the longitudinal direction, graphene ripples in the transverse direction to form ripples, creating a spatially varying strain profile. The periodic spatial variation of this strain modulates the C-C bonds in graphene, creating a superlattice of regions which are locally dense (turquoise, with short C-C bonds) and rare (yellow, with large C-C bonds) (Fig. 1c). Pseudopotentials arise from this strain profile as do associated pseudo-gauge fields, as illustrated in Fig. 1d. The pseudo-gauge fields, which are spatial derivatives of the potentials, become large in the presence of strain gradients and are hence maximized at the interfaces between dense and rare regions. From tight-binding calculations employing a sinusoidal strain profile, we find that $B_{max} \sim 100$ T and $E_{max} \sim 10^7$ V/m are required to match the experimental measurements. In analogy to traditional superlattices, where novel electronic states emerge at the interfaces of two different materials, quantized energy spectra arise due to the combined effects of interfacial oscillating pseudo (*B, E)* fields.

The values for $B_{max}$ and $E_{max}$ determined here from atomistic calculations may be underestimated as the experimentally determined ripple shape is found to better resemble a triangular waveform, characterized by sharp crests and troughs. This would indicate a much larger strain gradient in our system relative to a sinusoidal waveform used to estimate the deformation profile and therefore possibly much larger pseudo-gauge fields than those predicted by the model. We also note that the LDOS peaks in Fig. 3c do not line up exactly. This is likely because of small variations of the strain magnitude along the line on which spectra are taken. Along that line, the strain magnitudes as characterized by parameters $h(y)$ (out-of-plane displacement) and $u_y(y)$ (in-plane displacement) will vary. As discussed in the earlier, $h(y)$ and $u_y(y)$ are maximum at the crests and troughs of the ripples. It is difficult to take spectra along a line where both $h(y)$ and $u_y(y)$ parameters are absolutely constant, and hence some strain variation is observed in the spectra and their peak positions do not line up.

We also note that as rendered in Fig. 1d (by violet curved lines), alternating zones of oppositely directed pseudo-magnetic fields should also lead to the formation of oppositely propagating valley-Hall edge states[16,34,38,39]. These edge states can represent snake-like trajectories[40] at the interfaces between zones of opposite pseudo-magnetic fields[41–43]. Since the direction of these magnetic fields is valley dependent[16–18], our graphene ripples can also be a potential candidate for exploring valley dependent transport phenomena[9,41,44]. In our measurements, the spectral signatures of the snake states are seen in an increased intensity of the LDOS peaks at the ripple crests and troughs (Fig. 4a).

With our realization of pseudo-gauge fields which are simultaneously intense (~ 100 T) and modulated at short length scales (~ 1-10 nm), we can finally begin to realize theoretical proposals of valley filters and electron optics in graphene which require localized magnetic barriers[4–8,10–12]. It will also be interesting to explore the consequences of such strain profiles in a wide variety of 2D materials which should also support STREMS and where strain is known to significantly influence electronic properties[45–49]. The intensity of these inhomogeneous pseudo-fields should make realization of these proposals possible even at room temperatures[14].

**Funding:** This material is based upon work supported by the National Science Foundation under Grant No. 1229138. T.GN and M.T. acknowledge The Air Force Office of Scientific Research (AFOSR) grant 17RT0244. V.-H.N, A.L. and J.-C.C. acknowledge financial support from the the Francqui foundation and the F.R.S.-FNRS of Belgium through the research project (T.1077.15), from the Flag-Era JTC 2017 project "MECHANIC" (R.50.07.18.F), from the Fédération Wallonie-Bruxelles through the ARC on 3D nanoarchitecturing of 2D crystals (16/21-077), and from the European Union's Horizon 2020 research and innovation program (696656).

The Supporting Information is available free of charge on the ACS Publications website at DOI: xx.xxxx/acs.nanolett.xxxxxxx.

> Methods, details of the sample growth, negating other possible origins of the data, details of theoretical calculations, strain measurements, $E_n \propto n$ plots, Toy model for Dirac electrons in inhomogeneous magnetic fields.

**Author contributions:** R.B. conceived the project; R.B., L.P. built the custom instrument; T.G.N. prepared the samples; R.B., L.P. collected the data; R.B. performed analysis; R.B., V.H.N., A.L. performed theoretical modeling; all authors took part in interpreting the results; R.B., V.H.N., T.G.N., A.L., J.-C.C, M.T., E.W.H. wrote the paper; J.C.C., M.T., E.W.H. advised.

**Competing interests:** Authors declare no competing interests

# Supporting Information

## Strain Modulated Superlattices in Graphene

**Authors:** Riju Banerjee[1]*, Viet-Hung Nguyen[2], Tomotaroh Granzier-Nakajima[1], Lavish Pabbi[1], Aurelien Lherbier[2], Anna Ruth Binion[1], Jean-Christophe Charlier[2], Mauricio Terrones[1], Eric William Hudson[1]*

**Affiliations:**

[1]Department of Physics, The Pennsylvania State University, University Park, PA 16802, USA

[2]Institute of Condensed Matter and Nanosciences, Université catholique de Louvain, Chemin des étoiles 8, B-1348 Louvain-la-Neuve, Belgium

**\*Correspondence to:** ehudson@psu.edu, riju@psu.edu

# Methods

All data was obtained at 80 K in a custom built ultra-high vacuum (UHV) STM system using a SPECS Tyto head with cut Pt-Ir (80%-20%) tips. Part of the analysis was done using the software Gwyddion[50]. Though similar results have been observed with multiple tips on multiple samples, for consistency and calibration, all results presented here are obtained with a single tip on the same sample. Samples were transferred to the UHV environment within 10 minutes of growth to minimize air exposure. The sample was annealed at 300° C for about 1.5 hours in UHV to evaporate any adsorbent that might have settled on the surface during the transfer process. Similar observations have been made even after multiple annealing processes.

## S1. LPCVD growth and Raman characterization of graphene:

Growth was performed using low-pressure chemical vapor deposition (LPCVD). A piece of electropolished copper was placed in a quartz tube and positioned at the center of a high temperature furnace. A boat containing ammonia-borane was placed in the tube upstream of the furnace and wrapped with a heating belt. Subsequently the tube was pumped down to $10^{-2}$ Torr and flushed with 37.5 sccm $H_2$ and 212.5 sccm Ar gas for several minutes. Afterwards the gas was left flowing and the furnace temperature was raised to 1020°C at which point the heating belt was set for 50 °C. When the heating belt reached the set temperature, 10 sccm methane was flowed through the furnace for 5 min. Afterwards the methane was shut off and the furnace cooled naturally to room temperature. Some Boron and Nitrogen dopant atoms were seen embedded in the graphene lattice. They appear clearly in STM as bright spots due to their higher density of states. None were seen in regions where the strained graphene was observed.

Previous studies of graphene on Cu had observed it to be n-doped[51], but our Dirac point was near 0 meV (Fig. 3a). The dopants can shift the Dirac point[52]. Though no dopants were observed near the ripples we reported data on (and given the ease of seeing them in regions where we did find them, we are confident of their local absence) it seems that their presence leads to a global, rather than a local, doping effect. The arbitrariness of the Dirac point is usually an issue for correctly numbering the Landau level peaks in LDOS, but fortunately for us, we do not need to explicitly identify the Dirac point in our study as a spatially oscillating magnetic field results in equally spaced peaks.

Raman spectroscopy (Fig. S1) shows that the graphene is indeed monolayer with a characteristic 2D peak.

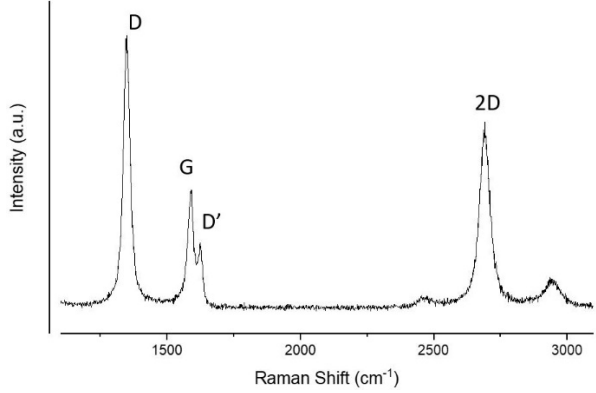

**Figure S1.** Raman spectrum ensures the 2D nature of the graphene sheet[53,54].

## S2. Negating other possible origins of LDOS peaks

Electrons in graphene can be confined by strain[27,28], and so it is important to consider whether the LDOS peaks we observe are effects of electrons trapped in a confining potential due to strain barriers created by ripple crests and troughs. Pseudo-electric barriers due to charge redistributions should be ineffective in confining electrons impinging perpendicularly on them due to Klein tunneling[55]. Pseudo-magnetic barriers, if strong enough can confine electrons in such a case[6], and such a hypothesis can be tested directly by taking spectra on regions with different ripple wavelengths. In such a case, we should expect the energy spacing between LDOS peaks to be reduced as the width of the confining potential is increased. In Fig. S2a, we plot the spectra taken on two regions with crest-to-trough distance of 14.0 nm (red) and 7.6 nm (black) taken on the same step edge. The energy spacing for both are almost same, in contrast to what would be expected from confinement effects[26].

Our model of periodic strain profile, on the other hand, demonstrates that the LDOS peak spacing is relatively insensitive to the wavelength. In general, energy spacing should be a function of wavelength $\lambda$ as well as out-of-plane $h_0$ and in-plane displacement $u_0$ parameters. Tight-binding calculations performed on rippled graphene show that the dependence on wavelength is indeed very weak (Fig. S2 b, c), as observed experimentally. Thus, while a quantization purely due to confinement creates states with energy spacing strongly dependent on the width of the well[26], strong spatially varying pseudo-magnetic fields create states with a different quantization.

Quasi-bound states due to a npn junction as observed by Bai et al[56] is another possible cause of the peaks. However, these states cannot be the result of a quasi-one dimensional quantization, as explained below. The step edges in our study are much taller than what they are in [56](our steps are several tens of nanometer high, which is an order of magnitude taller than the small facets observed there. That should significantly reduce the effects of any confinement in the vertical direction. As an example, we measured the spectrum on a step edge where there were no ripples (see Fig. S2 d, e below). Unlike the data presented in Fig 3 a, c of the main text, we do not see any peaks meaning that the effect of the confinement is indeed negligible in our case. The peaks we observe in the dI/dV spectra must then be the result of a pseudo-gauge field superlattice being formed by rippling of graphene.

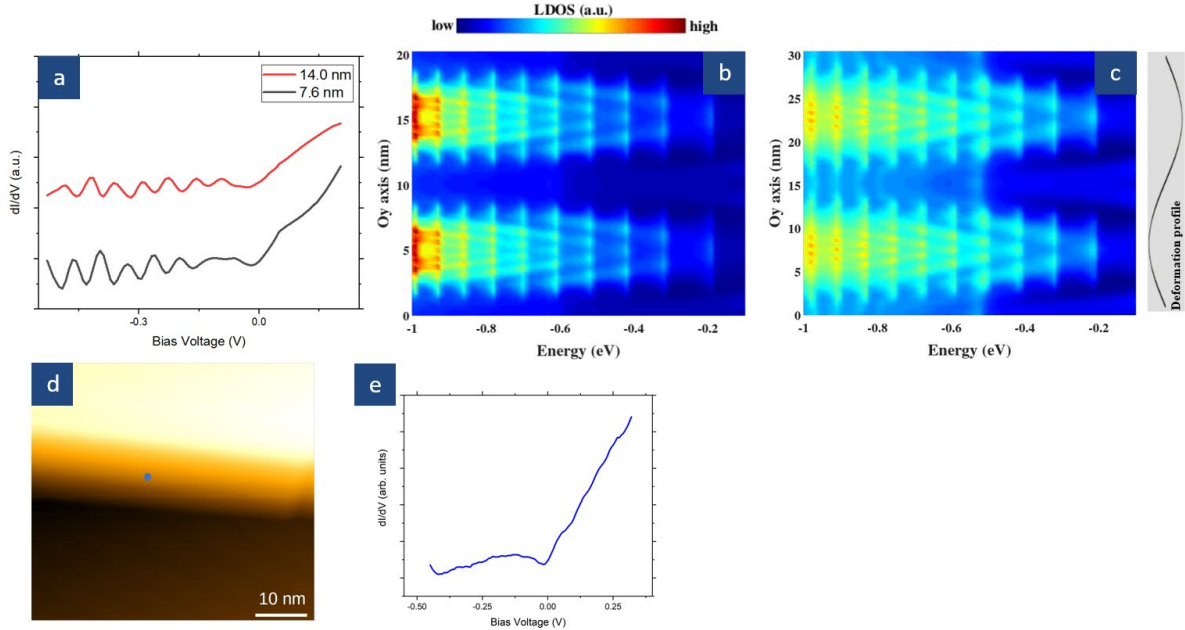

**Figure S2.** (a) Spectra taken on two regions on the same step edge with different ripple crest-to-trough distance (legend) show similar peak energies. If these peaks resulted from strain barrier-induced confinement effects, then the level energy spacing would vary significantly as the width of the confining potential varied. Spectra obtained with 13 mV bias modulation at ~971 Hz. Tight-binding calculations for our model, on the other hand, demonstrate similar energy spacings of LDOS peaks obtained for different ripple wavelengths of (b) 20 nm and (c) 30 nm. We measure the spectrum (e) on a region of a step edge where there are no ripples (blue dot in (d)) and we find that the equally spaced peaks we reported in Fig 3a, c of the main text to be missing. This proves that the peaks are not a result of any one-dimensional quasi-confinement of electrons but arise from a pseudo-gauge field superlattice being formed by periodic strain modulation. Our step edges are much taller than other ones reported in [56] so the effect of confinement is negligible.

## S3. Modeling methodologies: Tight binding versus DFT Calculations

Calculations based on tight-binding (TB) Hamiltonians have been demonstrated to be the most efficient approach to investigate the electronic properties of graphene systems, especially when their dimension reaches the nanometer regime. A tight-binding Hamiltonian adjusted due to lattice deformations [29] is able to compute accurately the effects of strain and to develop the theoretical models for analyzing the strain-induced pseudo-fields in graphene[19,21,41,44,57–59]. In particular, a first nearest-neighbor $p_z$-orbitals model, $H_{tb} = \sum_{\langle n,m \rangle} t_{nm} c_n^\dagger c_m$, with exponential dependence of the hopping energy $t_{nm}$ on the C-C bond length $r_{nm}$, $t_{nm} = t_0 \exp[-\beta(r_{nm}/r_0 - 1)]$, has been validated and widely employed to investigate the electronic properties of strained graphene without and even with small curvature[19,21,41]. In the rippled graphene systems considered here, a significant curvature is, however, obtained. Hence, the validity of the presented tight-binding model was re-examined, by fitting to first-principles calculations. In particular, Density Functional Theory (DFT) calculations implemented in the SIESTA package were performed to compute the electronic band-structure of reasonably small rippled systems and, accordingly, the hopping parameters of the TB Hamiltonian were properly adjusted to agree with the obtained DFT data.

The band-structures of rippled graphene systems with $\lambda = 10$ nm, $h_0 = 1.0$ nm and $u_0 = 2.0$ Å obtained by both DFT and our adjusted TB calculations are presented in Fig. S3. To get the best fit between these two calculations, the decay rate $\beta$ of the TB Hamiltonian has to be adjusted to ~ 4.5 while a value of 3.37 is usually used in flat graphene systems with in-plane deformations only[29]. This value of $\beta \approx 4.5$ is valid in the cases of extreme strains (~ 10%) considered in this work, which is consistent with direct strain measurements presented in

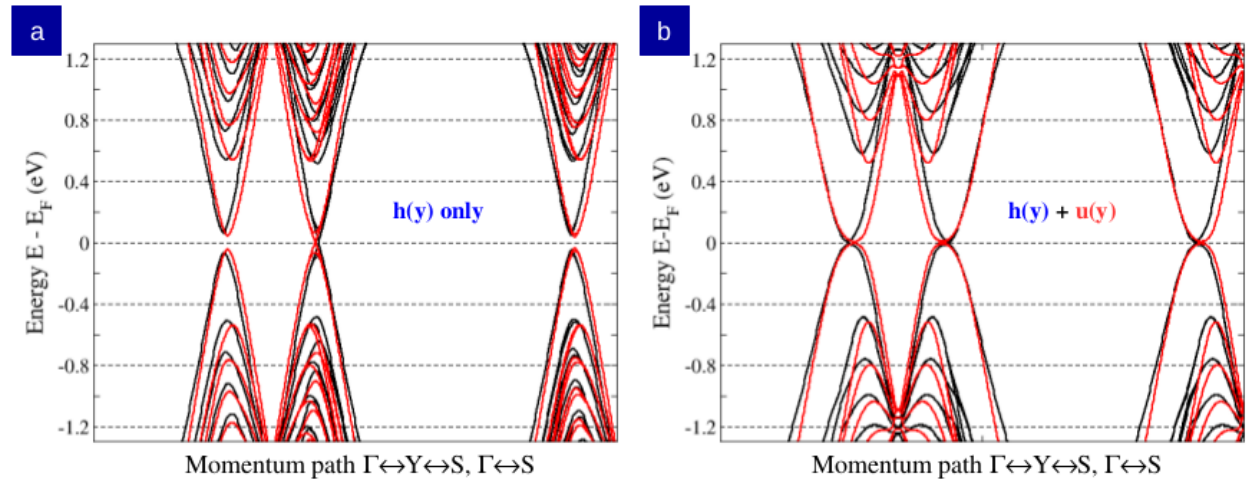

**Figure S3**. Band structure of rippled graphene systems (a) without and (b) with in-plane displacements: tight binding approach (red curves) versus DFT calculations (black curves). Ripple parameters $\lambda = 10$ nm, $h_0 = 1.0$ nm and $u_0 = 2.0$ Å are considered.

Supplementary sec. 5. The validity of this adjusted TB Hamiltonian is further confirmed by the LDOS maps presented in Figs. S4(a, b). This TB Hamiltonian was finally employed to investigate the electronic spectra of rippled graphene systems, presented in the main text, with similar sizes as in the experiments.

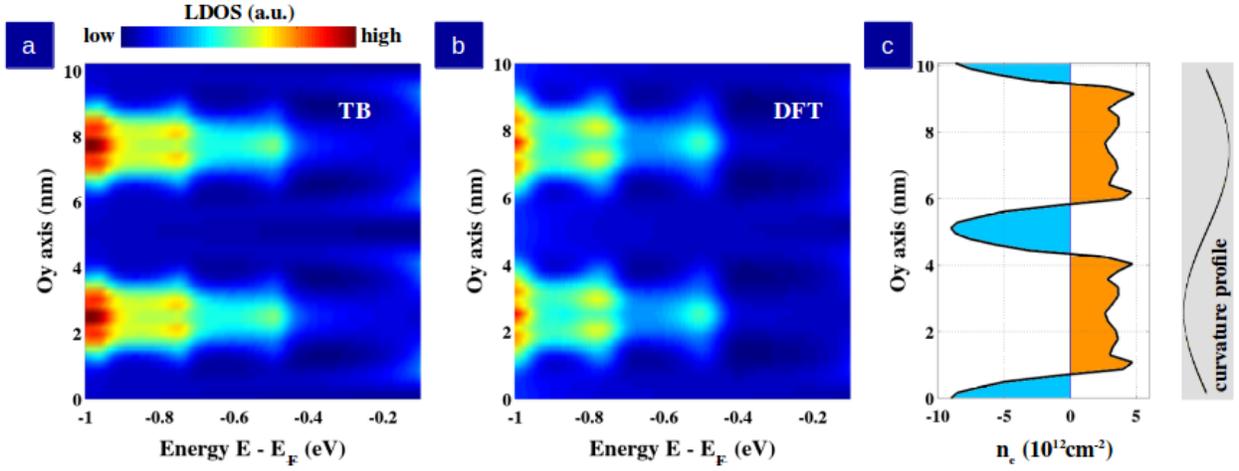

**Figure S4**. LDOS maps obtained by (a) tight-binding and (b) DFT calculations. (c) Local charge density along the ripple axis obtained by DFT calculations. The rippled graphene system of Fig.S3b is considered here.

The pseudo-vector $\vec{A}$ and scalar $V_S$ potentials in the effective model can be determined in the continuous approximation as a function of the strain tensor $\epsilon_{ij}(\vec{r})$ [16–18]:

$$A_x - iA_y = \pm \frac{\hbar \beta}{2er_0}\left(\epsilon_{xx} - \epsilon_{yy} + 2i\epsilon_{xy}\right) \quad (2)$$

$$V_S = \frac{\mu_0 r_0}{2}\left(\epsilon_{xx} + \epsilon_{yy}\right) \quad (3)$$

where $\epsilon_{ij} \equiv \frac{1}{2}\left(\frac{\partial u_i}{\partial r_j} + \frac{\partial u_j}{\partial r_i}\right) + \frac{1}{2}\frac{\partial h}{\partial r_i}\frac{\partial h}{\partial r_j}$ and $\mu_0$ is the characteristic energy function [60]. The ± signs in Eq. (2) correspond to the vector potential applied to fermions in the *K/K'* valleys of the graphene Brillouin zone, respectively. In our simple strain profile, Eqs. (2) and (3) reduce to $A_x = \mp \frac{\hbar \beta}{2er_0}\epsilon_{yy}$ and $V_S = \frac{\mu_0 r_0}{2}\epsilon_{yy}$ respectively.

In addition, *ab initio* calculations were also used to compute the local charge density (Fig. S4c) along the ripple of the system studied in Fig. S3b. It is shown that due to its lattice (and hence electronic) inhomogeneity, charge carriers are redistributed, and a non-uniform local charge density is observed along the ripple axis. This result essentially explains the necessity of taking into account both pseudo-magnetic and electric fields in the effective calculations, as presented in the main text, to accurately depict the electronic properties of the system.

## S4. Observation of pseudo-electric fields

Note that even though both pseudo-magnetic and electric fields can be induced by a strain field, it has been shown that the electric field is generally less pronounced [20,58–60], because of screening effects. As discussed in the previous section, a doping inhomogeneity is actually obtained along the ripple axis, thus implying that the pseudo-electric fields also occur and can play an important role on the electronic properties of the considered rippled systems. This is demonstrated in the main text by comparison between calculations using a strained TB Hamiltonian and an unstrained one with pseudo-fields and is further confirmed by the analysis below.

First, if only the pseudo-magnetic field as in the main text is considered, there are two zones of positive fields interleaved with two other zones of negative fields in each periodic unit of rippled graphene. Under the effect of such magnetic fields, opposite edge states are formed at the transition of opposite fields, and, accordingly, the LDOS is a periodic function along the ripple axis with a periodic length $\lambda/4$ as illustrated in Fig.S5a. However, this is not the case observed in experiments and confirmed by calculations using the strained tight-binding Hamiltonian, where a periodic length of $\lambda/2$ is obtained. Such a picture can be however obtained when both pseudo-magnetic field and electric fields are considered (Fig. 4b of main text). When a pseudo electric field (i.e., periodic potential energy as considered) is added, the spatial variation of the potential energy gives rise to another feature that redistributes electrons, leading to much more available hole edge states in the local zone of high potential energies whereas the number of hole edge states in the local zone of low potential energies are reduced (see Fig. S4b). The existence of this electric field and accordingly the corresponding charge redistribution (i.e., inhomogeneous distribution of charges) is further confirmed by *DFT* calculations (see Fig. S4c).

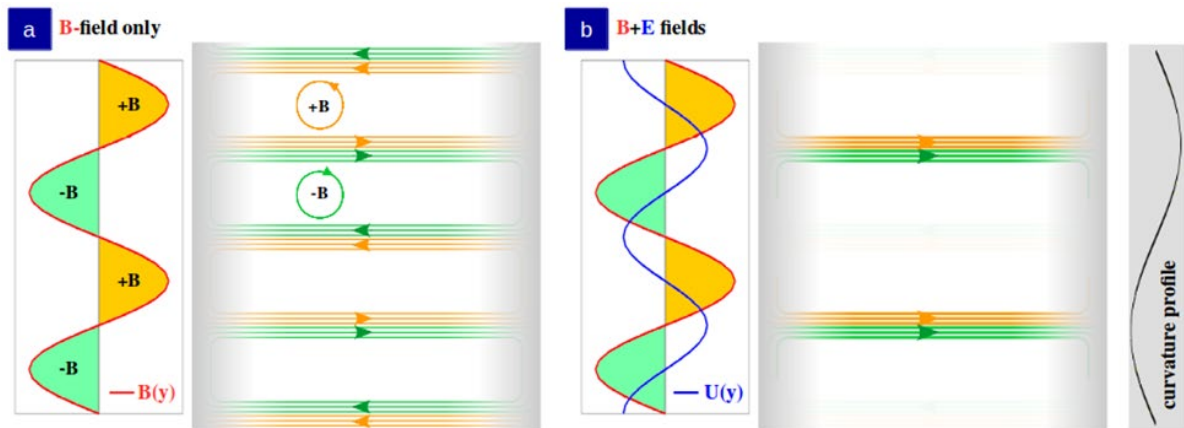

**Figure S5**. Schematics illustrating the effects of spatially oscillating pseudo- (E, B) fields.

## S5: Extreme strain measurement

It is difficult to directly image the strain variation over a ripple by measuring individual atomic displacements. The fact that C-C bond lengths are larger at the crests and troughs of the ripples and smaller in the region between them was deduced by fitting our experimental observations (equally spaced LDOS peaks with higher peal amplitude at the ripple crests and troughs) to a theoretical model (Fig. 4). However, due to the triangular shape of ripples (shown in Fig. 4a), the strain magnitude varies predominantly only around a very narrow region near the crests and troughs. Hence, we can estimate the longitudinal strain in our system by imaging the lattice in a sufficiently small region lying between the ripple crests and troughs. We estimate strain in our system by comparing Fourier transforms of atomic resolution images taken on unstrained and strained regions (Fig. S6 a, b respectively). As expected, the unstrained graphene lattice is characterized by hexagonal periodicity with Fourier space peaks (marked by white dashed circles) equidistant from the center, falling on a centered (grey dotted) circle. On the other hand, graphene imaged on the draped region yields a clearly distorted hexagonal lattice. The deviation in magnitudes of the wavevectors from that of unstrained graphene is used to estimate the strain. Ratios of lattice constants imply strain magnitudes over 10%. This measurement is also consistent with first principle calculations required to reproduce the observed spectra, as discussed in Supplementary sec. 3. Both regions are imaged with the same tip to reduce potential calibration issues. The two regions are also chosen to be close to each other to eliminate any potential effects of non-linearity in piezo response.

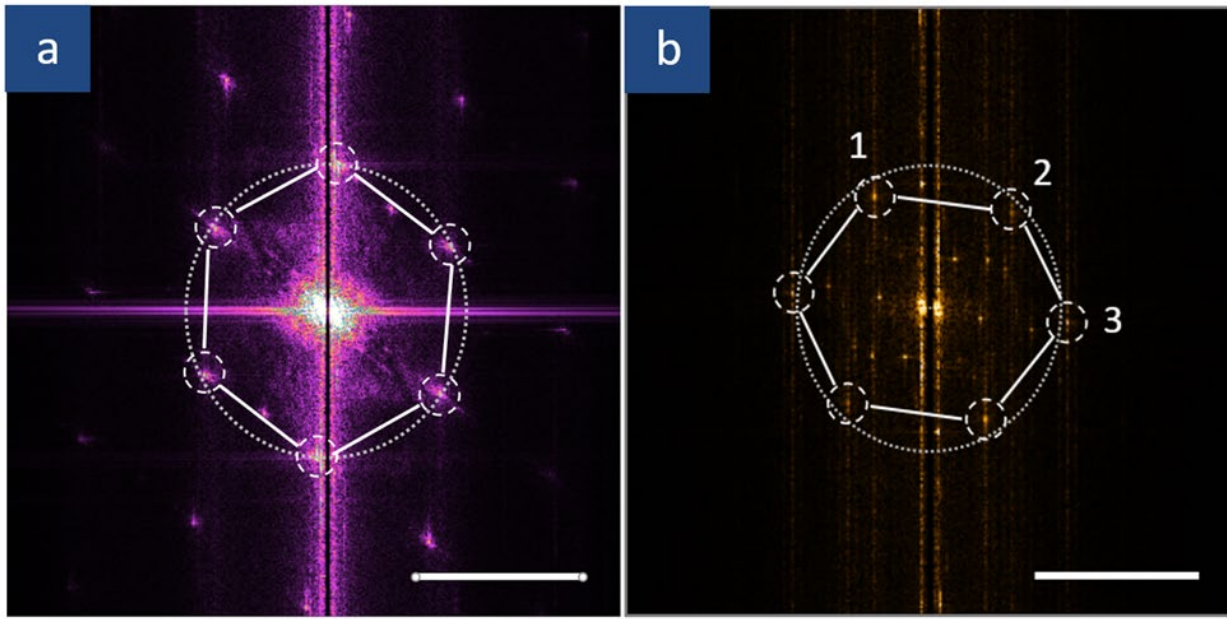

**Figure S6.** Direct measurement of extreme strain: (a) FFT of atomic resolution topography from an unstrained graphene lattice. Peaks (marked by white dashed circles for clarity) are equidistant from the center (as grey dotted circle). (b) Topographic FFT from the draped region shows lattice distortion due to strain. Measuring deviation from the unstrained positions (grey circle) yields strains of 1) 10.3%, 2) 6.8% and 3) -3.3%. Scalebars in both figures are 5 nm$^{-1}$ (= 0.2 nm).

Previous studies[61] have reported graphene sheets sticking to the STM tips. Such effects are particularly large for micron sized graphene sheets which are free to bend and should not be important for graphene ripples measuring only tens of nanometers and stretched taut under tension. Similar topographic images were obtained with forward and backward motion of the STM tip, proving that the graphene position is not appreciably disturbed by the tip. The forward and backward scans also yielded topographies showing similar strain magnitudes. Increasing the tunneling current by an order of magnitude (~100 pA to ~1 nA) did not change the topography, further proving that tip artifacts are negligible in the topography.

We also note that in making measurements from topographies of the draped region, the typical approach to STM analysis of using plane subtraction will, in such a highly sloped region, lead to artificial compression along the draping direction (it is equivalent to projection into a plane parallel to the terraces). Thus, we instead rotate the coordinate system in order to properly extract all distances in the draped graphene.

## S6: $E_n \propto n$ plots

In Fig. 2c we plotted the background subtracted spectra taken on a STREMS and plotted the energy dependence of the peaks for Spectrum #19 in Fig. 2d. In Fig. S7, we show the linear fits for all twelve spectra taken on the draped part (numbered #8-19) with their linear fits. We observe that the higher energy peaks deviate more from the linear fits. This can be attributed to the fact that graphene dispersion deviates from being linear at higher energies away from the Dirac point.

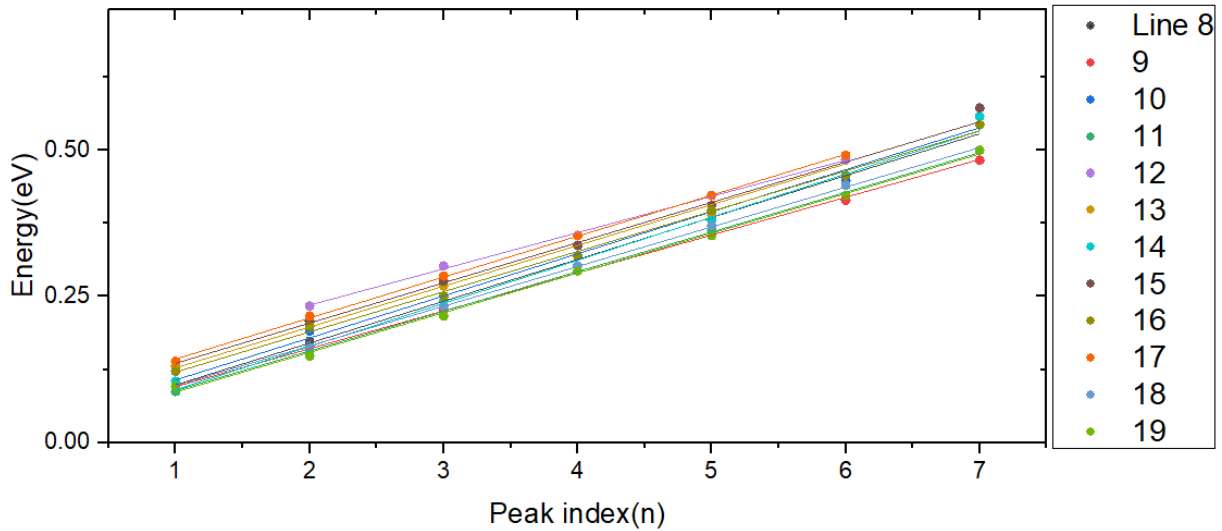

**Figure S7.** All twelve background subtracted spectra (Line #8-19) taken on the STREMS and shown in Fig. 2c of the text have equally spaced peaks.

**S7: A toy model showing $E_n \propto n$ quantization of Dirac electrons by a periodically oscillating magnetic field profile**

Fig 5a, b of the main text shows simulations where a periodically varying magnetic field profile results in $E_n \propto n$ DOS peaks as the periodicity is progressively reduced and made comparable to the magnetic length. In this supplementary section, we derive the same result analytically by considering a simple spatially oscillating magnetic field profile.

The Hamiltonian for Dirac electrons in presence of a magnetic field is given by

$$H = v_F \vec{\sigma} \cdot (\vec{p} + e\vec{A})$$

To demonstrate the case of a periodically oscillating field, we consider here the simplest case: where the B fields are a series of alternating $\delta$-functions separated by distance L (Fig. S8). As $\vec{B} = \vec{\nabla} \times \vec{A}$, in 1D, $A = constant$. Let's assume $\vec{A} = (0, A, 0)$.

Then the Hamiltonian in this simple case becomes:

$$H = v_F[\sigma_x p_x + \sigma_y A] = v_F \begin{bmatrix} 0 & p_x - iA \\ p_x + iA & 0 \end{bmatrix}$$

This Hamiltonian must satisfy $H\psi = E\psi$.

$$v_F \begin{bmatrix} 0 & p_x - iA \\ p_x + iA & 0 \end{bmatrix} \begin{pmatrix} \psi_1 \\ \psi_2 \end{pmatrix} = E \begin{pmatrix} \psi_1 \\ \psi_2 \end{pmatrix}$$

Which gives

$$\frac{d\psi_2}{dx} + A\psi_2 = i\xi\psi_1$$

And

$$\frac{d\psi_1}{dx} - A\psi_1 = i\xi\psi_2$$

Where we define $\xi \equiv \frac{E}{v_F}$

These two coupled equations can be decoupled by making a second order differential equation:

$$\frac{d^2\psi_1}{dx^2} + (\xi^2 - A^2)\psi_1 = 0$$

And the same for $\psi_2$. So each spinor component $\psi_{1,2}$ satisfies a second order differential equation like Schrodinger equation of massive particles, with an effective potential of $V_{eff} = -A^2$ and effective energy $\xi^2 = (E/v_F)^2$. Next, we call $\psi_{1,2}$ as $\varphi$ as they both satisfy the same differential equation and solve it in the two regions marked by I and II in Fig. S8.

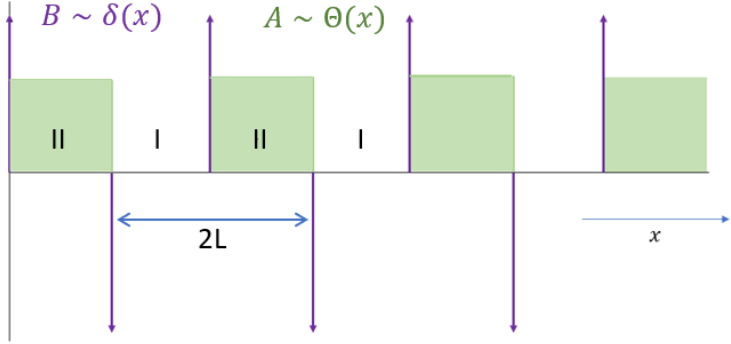

**Figure S8:** Modeling the effect of periodic magnetic field profile on Dirac electrons. We choose a simple case where all the magnetic fields are alternating delta functions and the vector potential are step functions. As our ripples have a triangular shape, most of the strain gradient is concentrated at a small region near the crest and troughs of the ripples, making the delta function approximation of the pseudo-magnetic field valid.

$\varphi_I$ satisfies the equation:

$$\frac{d^2 \varphi_I}{dx^2} = -\xi^2 \varphi_I$$

Solution should be of the form: $\varphi_I = A_r e^{i\xi x} + A_l e^{-i\xi x}$

And similarly, $\varphi_{II}$ should have the form:

$\varphi_{II} = B_r e^{ik_1 x} + B_l e^{-ik_1 x}$ where $k_1 \equiv \sqrt{\xi^2 - A^2}$ when $\xi^2 > A^2$

Or $\varphi_{II} = B_r e^{ik_2 x} + B_l e^{-ik_2 x}$ where $k_2 \equiv \sqrt{A^2 - \xi^2}$ when $A^2 > \xi^2$.

Focusing on scattering states $\xi > A$, we enforce continuity at $x = 0$ to get:

$$A_r + A_l = B_r + B_l \qquad \qquad i$$

$$i\xi(A_r - A_l) = ik_1 B_r - ik_1 B_l \qquad \qquad ii$$

For a periodic structure, Bloch theorem ensures:

$$\varphi_I(0) = \varphi_{II}(2L) e^{ik(2L)} \qquad \qquad iii$$

And

$$\varphi_I'(0) = \varphi_{II}'(2L) e^{ik(2L)} \qquad \qquad iv$$

For these four equations (*i, ii, iii, iv*) to have a solution, the determinant of the coefficients must be zero. That determines the coefficients $A_r, A_l, B_r, B_l$.

$$D = \begin{bmatrix} 1 & 1 & -1 & -1 \\ i\xi & -i\xi & -ik_1 & ik_1 \\ 1 & 1 & -e^{i2L(k+k_1)} & -e^{i2L(k-k_1)} \\ i\xi & i\xi & -ik_1 e^{i2L(k_1+k)} & +ik_1 e^{i2L(k-k_1)} \end{bmatrix}$$

The condition det(D) = 0 gives us:

$$(-1 + e^{i2L(k-k_1)})(-1 + e^{i2L(k+k_1)})\xi k_1 = 0 \qquad v$$

The above expression is satisfied when either of the two terms within brackets is zero.

For the expression in the left bracket, this means $e^{i2L(k-k_1)} = 1$. This implies $cos(2L(k - k_1)) = 1$ and $sin(2L(k - k_1)) = 0$. Both these conditions are satisfied when $2L(k - k_1) = \pm n\pi$. This gives an expression for the energy bands: $\xi^2 = \left(k \pm \frac{n\pi}{2L}\right)^2 + A^2$, where n is an integer. As we assumed $A$ to be a constant, we get can drop that as a constant energy shift and end up with:

$$\xi = \left(k \pm \frac{n\pi}{2L}\right)$$

We obtain an identical expression for energy bands when considering the other term within brackets in equation *v*. Thus, the bandstructure consists of a series of Dirac cones separated by $\frac{n\pi}{2L}$, creating a DOS with equally spaced peaks.